\documentstyle[aps,floats]{revtex}
\newcommand{\ud}{d}
\newcommand{\damnbox}{\Box}
\begin{document}
\twocolumn[\hsize\textwidth\columnwidth\hsize\csname 
           @twocolumnfalse\endcsname
\title{Radiative falloff in black-hole spacetimes} 
\author{William G.~Laarakkers and Eric Poisson} 
\address{Department of Physics, University of Guelph, Guelph,
         Ontario, Canada N1G 2W1}
\maketitle
\begin{abstract}
This two-part contribution to the Proceedings of the Eighth Canadian
Conference on General Relativity and Relativistic Astrophysics is
devoted to the evolution of a massless scalar field in two black-hole
spacetimes which are not asymptotically flat.  

In Part I (authored by Eric Poisson) we consider the evolution of a
scalar field propagating in Schwarzschild-de Sitter spacetime. The
spacetime possesses a cosmological horizon in addition to the usual
event horizon. The presence of this new horizon affects the late-time
evolution of the scalar field.    

In part II (authored by William G.~Laarakkers) we consider the
evolution of a scalar field propagating in Schwarzschild-Einstein-de 
Sitter spacetime. The spacetime has two distinct regions: an inner
black-hole region and an outer cosmological region. Early on in the
evolution, the field behaves as if it were in pure Schwarzschild
spacetime. Later, the field learns of the existence of the
cosmological region and alters its behaviour. 
\end{abstract}
\vskip 2pc]

\narrowtext

\begin{center}
\large \bf
Part I \\
by Eric Poisson
\end{center}

\subsection*{Introduction} 

A theorem that establishes the uniqueness of the Schwarzschild black
hole as the endpoint of gravitational collapse without rotation was
proved by Werner Israel more than 30 years ago \cite{1}, and the
mechanism by which the gravitational field eventually relaxes to the
Schwarzschild form was elucidated by Richard Price more than 25 years
ago \cite{2}. Given the venerable age of this topic, it is surprising
that more can be said about it today. Yet, many papers on radiative
falloff have been written  in the last few  
years \cite{3,4,5,6,7,8,9,10,11,12,13,14,15}. Most of the new
developments are concerned with rotating collapse, and how the
gravitational field eventually relaxes to the Kerr form. The
question we pursue in this two-part contribution is
different. Focusing our attention on nonrotating black holes, we ask:
How do the conditions far away from the black hole affect the
relaxation process? In Part I we consider a black hole immersed in an
inflationary universe. (This was first done by Brady 
{\it et al.}~\cite{14}, and additional details can be found in
Ref.~\cite{15}.) In Part II, William G.~Laarakkers will consider a
black hole immersed in a spatially-flat, dust-filled universe.    

\subsection*{Radiative falloff in Schwarzschild spacetime}

Price's result \cite{2} can be summarized as follows. As a
nonspherical star undergoes gravitational collapse, the gravitational
field becomes highly dynamical, and the escaping radiation interacts 
with the spacetime curvature surrounding the star. At late times, well
after the initial burst of radiation was emitted, the gravitational
field relaxes to a pure spherical state. If $\delta g$ schematically
represents the deviation of the metric from the Schwarzschild
form, then $\delta g \sim t^{-(2\ell + 2)}$, where $\ell$ is the
multipole order of the perturbation; the dominant contribution to
$\delta g$ comes from the quadrupole ($\ell = 2$) mode.  

The inverse power-law decay applies to many other situations involving 
radiation interacting with the curvature created by a massive
object. The simplest model problem which exhibits this behaviour
involves a massless scalar field in Schwarzschild spacetime. In this
context, the background geometry is not affected by the field $\Phi$,
which satisfies the wave equation
\begin{equation}
\bigl( g^{\alpha\beta} \nabla_{\!\alpha} \nabla_{\!\beta} 
- \xi R \bigr) \Phi = 0,
\label{1}
\end{equation}
where $g_{\alpha\beta}$ is the spacetime metric, $R$ the Ricci scalar
(which vanishes for Schwarzschild spacetime), and $\xi$ a coupling
constant. Because the spacetime is spherically symmetric, the field
can be decomposed according to 
\begin{equation}
\Phi = \sum_{\ell m} \frac{1}{r}\, \psi_\ell(t,r)\, 
Y_{\ell m}(\theta,\phi).
\label{2}
\end{equation}
This leads to a decoupled equation for each wave function
$\psi_\ell$, and we can focus on a single mode at a time. 

The problem is formulated as follows. A pulse of scalar radiation
(described by $\psi_\ell$) impinges on the black hole and interacts
with the spacetime curvature, which creates a potential barrier fairly
well localized near $r=3M$. The wave pulse is partially reflected and 
transmitted, and at late times, a tail remains. At such times, the
field falls off as $\psi_\ell \sim t^{-(2\ell + 3)}$. This is Price's
power-law decay, and this behaviour is displayed in Fig.~1.     

A number of analytical and numerical studies of radiative dynamics
\cite{3,4,5,6,7} have revealed that the inverse power-law behaviour is
not sensitive to the presence of an event horizon. In fact, power-law
tails are a weak-curvature phenomenon, and it is the asymptotic
structure of the spacetime at radii $r \gg 2M$ which dictates how the
field behaves at times $t \gg 2M$. It is this observation that
motivated our work: How is the field's evolution affected if the
conditions at infinity are altered?  

\begin{figure}
\special{hscale=35 vscale=35 hoffset=-25.0 voffset=10.0
         angle=-90.0 psfile=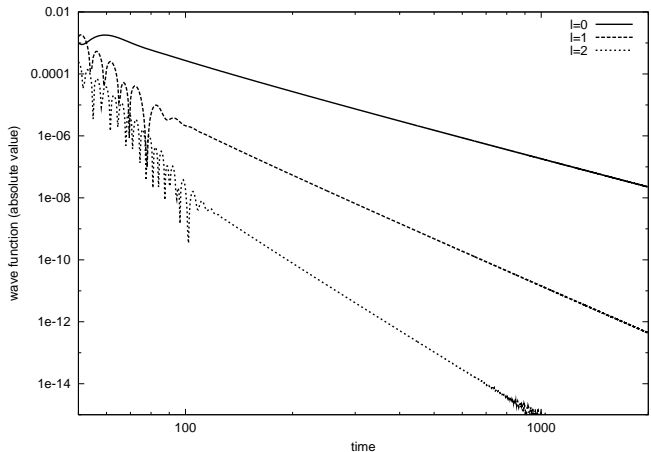}
\vspace*{2.6in}
\caption{Absolute value of the wave function $\psi_\ell(t,r)$ as a
function of time $t$, evaluated at $r = 10$ in Schwarzschild
spacetime. We use units such that $2M = 1$. The cases $\ell=0,1,2$ are 
considered, and the wave functions are plotted on a log-log scale. In
such a plot, a straight line indicates power-law behaviour, and a
change of sign in the wave function is represented by a deep
trough. We see that the field's early behaviour is oscillatory,
but that it eventually decays according to an inverse power law.}  
\end{figure}

\subsection*{Radiative falloff in Schwarzschild-de Sitter spacetime} 

To provide an answer to this question, we remove the black hole from
its underlying flat spacetime and place it in de Sitter spacetime,
which describes an exponentially expanding universe. The
Schwarzschild-de Sitter (SdS) spacetime has a metric given by 
\begin{eqnarray}
ds^2 &=& -f\, dt^2 + f^{-1}\, dr^2 + r^2\, d\Omega^2, 
\nonumber \\ & & \label{3} \\
f &=& 1 - 2M/r - r^2/a^2. \nonumber
\end{eqnarray}
Here, $a^2 = 3/\Lambda$, where $\Lambda$ is the cosmological
constant. (The SdS metric is a solution to the modified vacuum field
equations, $G_{\alpha\beta} + \Lambda g_{\alpha\beta} = 0$, which
imply $R = 4\Lambda = 12/a^2$.) The spacetime possesses an event
horizon at $r = r_e \simeq 2M$ and a cosmological horizon at $r = r_c
\simeq a$. We assume that $r_e \ll r_c$, so that the two length scales
are cleanly separated. 

\begin{figure}
\special{hscale=35 vscale=35 hoffset=-25.0 voffset=10.0
         angle=-90.0 psfile=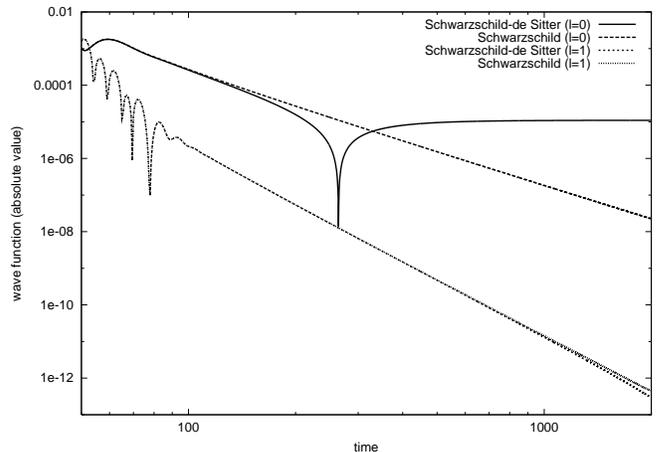}
\vspace*{2.6in}
\caption{Absolute value of the wave function $\psi_\ell(t,r)$ as a
function of time $t$, evaluated at $r = 10$ in Schwarzschild
spacetime ($r_e = 1$) and SdS spacetime ($r_e = 1$ and $r_c =
2000$). The cases $\ell=0, 1$ are considered, and the wave functions
are plotted on a log-log scale. In both cases, $\xi = 0$.}   
\end{figure}

We examine the time evolution of a scalar field in SdS spacetime; the 
field is still governed by Eq.~(\ref{1}), and it still admits the
decomposition of Eq.~(\ref{2}). Figure 2 provides a comparison between
the behaviour of $\psi_\ell$ in the two spacetimes (Schwarzschild and
SdS). We see that at early times, the wave functions behave
identically; the field has not yet become aware of the different
conditions at $r \gg r_e$. At later times, however, deviations become
apparent. For $\ell=0$, the Schwarzschild behaviour $\psi_0 \sim
t^{-3}$ is replaced by the wave function changing sign at $t \sim
260$, and settling down to a constant value at late times. For
$\ell=1$, the Schwarzschild behaviour $\psi_1 \sim t^{-5}$ is replaced
by a faster decay which eventually becomes exponential.

The field's exponential decay is confirmed by monitoring its evolution
up to times $t > r_c$. If $\xi = 0$, we find that $\psi_\ell \sim
e^{-\ell \kappa_c t}$ at late times \cite{14}, where $\kappa_c \simeq
1/r_c$ is the surface gravity of the cosmological horizon. 

A rich spectrum of late-time behaviours is revealed when $\xi$, the
curvature-coupling constant, is allowed to be nonzero. Figure 3
shows the time-dependence of $\psi_0$ for several values of $\xi$. 
For $\xi$ smaller than a critical value $\xi_c$, the field decays 
monotonically with a decay constant that increases with increasing
$\xi$. When $\xi > \xi_c$, however, the wave function oscillates
with a decaying amplitude. As $\xi$ is increased away from the
critical value $\xi_c$, the frequency of the oscillations increases,
but the decay constant stays the same. 

\begin{figure}
\special{hscale=35 vscale=35 hoffset=-25.0 voffset=10.0
         angle=-90.0 psfile=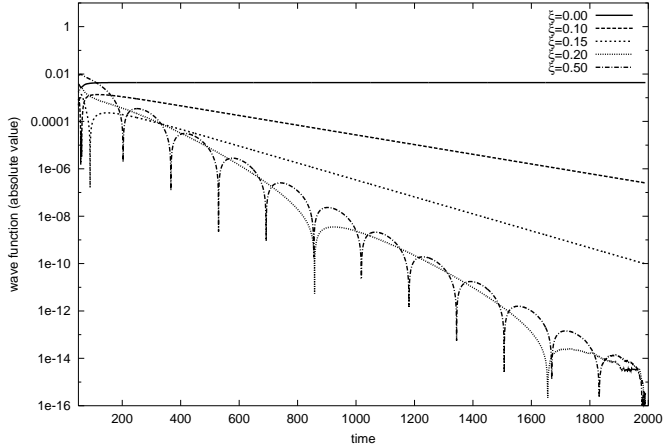}
\vspace*{2.6in}
\caption{Absolute value of the wave function $\psi_0(t,r)$ as a
function of time $t$, evaluated at $r = 10$ in SdS spacetime 
($r_e = 1$ and $r_c = 100$). Several values of $\xi$ are considered,
in the interval between $\xi = 0$ and $\xi = \frac{1}{2}$. The wave
functions are plotted on a semi-log scale, in which a straight line
indicates exponential behaviour.} 
\end{figure}

This qualitative change of behaviour as $\xi$ goes through 
$\xi_c$ is quite remarkable. It can be explained with a detailed
analytical calculation that will not be presented here (see
Ref.~\cite{15}). This calculation reveals that at late times, the
field behaves as $\psi_\ell \sim e^{-p \kappa_c t}$, where
\begin{equation}
p = \ell + \frac{3}{2} - \frac{1}{2}\sqrt{9-16\xi} 
+ O\biggl ( \frac{r_e}{r_c} \biggr).
\label{4}
\end{equation}
This relation implies that $p$ becomes complex, and $\psi_\ell$
oscillatory, when $\xi > \xi_c \equiv 3/16$.   

\begin{center}
\large \bf
Part II \\
by William G.~Laarakkers
\end{center}

\subsection*{The spacetime}

\begin{figure}
\special{hscale=60 vscale=60 hoffset=35.0 voffset=-180.0
         angle=0.0 psfile=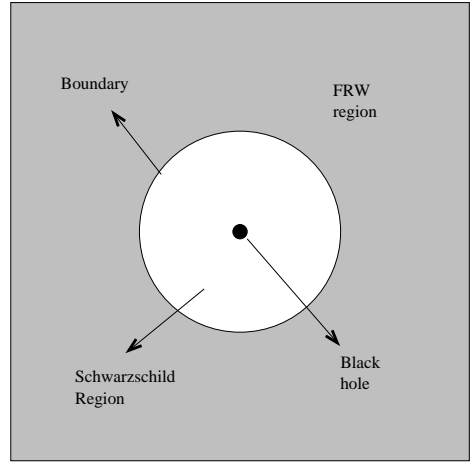}
\vspace*{2.6in}
\caption{Schematic of the Schwarzschild-Einstein-de Sitter
spacetime. Note that the boundary between the two regions is expanding
outwards, co-moving with the universe.} 
\label{fig:space} 
\end{figure}

The background spacetime in which the scalar field's evolution is
followed is the Schwarzschild-Einstein-de Sitter spacetime.
Qualitatively, it can be described as follows.  The idea is to start
out with a spatially-flat, expanding, dust-filled universe. Then a
ball of dust is ``scooped'' out, which leaves behind a spherical
vacuum region. The dust that was removed is replaced by a
Schwarzschild black hole, which is placed in the middle of the vacuous
region. This produces a spacetime with two distinct regions. The inner
(black hole) region is described by the Schwarzschild metric, and the
outer (cosmological) region is described by the
Friedman-Robertson-Walker (FRW) metric (see Fig.~\ref{fig:space}). 

There are two important things to note about this spacetime.  First,
if the mass of the black hole is the same as the mass of the dust that
was scooped out, the metric will be smooth across the boundary
separating the two regions of the spacetime.  Also, since the dust is
pressureless it will not flow across the boundary, and the boundary
itself will be co-moving with the universe. 
 
Because the specific finite-difference equation used in the numerical
work requires the use of null coordinates (see \cite{3}), the
metrics of the two regions must be put in double-null coordinate form.
For the black hole region the metric is written as 
\begin{eqnarray}
\ud s^2 = -\left(1-\frac {2M}{r}\right)\, \ud u \ud v\, + 
r^2\, d\Omega^2.
\end{eqnarray}
Here, $u$ and $v$ are ingoing and outgoing null coordinates, and $r$
is defined implicitly by $r+2M \ln(r/2M-1)=(v-u)/2$.  In the
cosmological region the metric takes the form
\begin{eqnarray}
\ud s^2 = a^2(u^*,v^*)\hspace{1mm}(-\ud u^* \ud v^* 
+ \chi ^2\, d\Omega^2),
\end{eqnarray}
where $u^*$ and $v^*$ are ingoing and outgoing null coordinates of the
FRW spacetime, different from $u$ and $v$.  The FRW radial coordinate
is $\chi = \frac{1}{2} (v^*-u^*)$. The scale factor $a$ is given by 
$a(u^*,v^*) = \frac{1}{16}C(u^*+v^*)^2$, where $C$ is a constant that
depends on the mass $M$ of the black hole and the density of the dust.
 
The first task is to find one coordinate system that can describe both
regions of the spacetime.  This is required so that a single wave
equation valid over the entire spacetime can be constructed.  Since it
is known that the metric is continuous across the boundary we can
evaluate the metric induced on both sides of the boundary
hypersurface, and set them equal.  This construction allows us to find
the ingoing Schwarzschild coordinate $u$ as a function of the ingoing
cosmological coordinate $u^*$, and the outgoing coordinate $v$ as a
function of $v^*$.  Thus we now have a single coordinate system
covering both regions of the spacetime.

\subsection*{The wave equation}

The wave equation that governs the evolution of the scalar field is
one without curvature coupling (equivalent to setting $\xi = 0$ (see
Part I and \cite{15}). Thus the massless scalar field $\Phi$
obeys the equation
\begin{eqnarray}
\damnbox \Phi = g^{\alpha \beta}\nabla_\alpha \nabla_\beta \Phi= 0.
\end{eqnarray}
The spherical symmetry of the problem allows us to decompose the field
in terms of spherical harmonics, and then to evolve only the part of
the field that depends on the null coordinates.  Thus the field can be
decomposed as
\begin{eqnarray}
\Phi = \sum_{\ell m} \frac{1}{R}\, \psi_\ell\, Y_{\ell m}(\theta, \phi), 
\end{eqnarray}
where $R=r$, $\psi_\ell = \psi_\ell(u,v)$ in the black hole region,
and $R=a\chi$, $\psi_\ell = \psi_\ell(u^*,v^*)$ in the cosmological
region.  When all quantities are expressed in the starred coordinate
system, each wavefunction $\psi_\ell$ satisfies the equation
\begin{eqnarray}
4\, \frac {\partial{^2\psi}}{\partial{u^*} \partial{v^*}}
\hspace{1mm}+\hspace{1mm}V \, \psi = 0,
\end{eqnarray}
where the potential $V$ takes a different form depending on which
region of the spacetime the field lies:
\begin{eqnarray}
V_{\rm Schild}&=&\frac{\ud u}{\ud u^*} \frac{\ud v} {\ud v^*} 
f \left[ \frac {\ell(\ell+1)} {r^2} + \frac {2M} {r}\right] 
\eqnum{10$a$} \\
\vspace{1cm}
V_{\rm FRW}&=&\frac{4\,\ell(\ell+1)}{(v^*-u^*)^2} 
- \frac {8} {(v^*+u^*)^2}. 
\eqnum{10$b$}
\label{eq:VFRW}
\end{eqnarray}

\subsection*{Results}

\begin{figure}
\special{hscale=35 vscale=35 hoffset=-25.0 voffset=10.0
         angle=-90.0 psfile=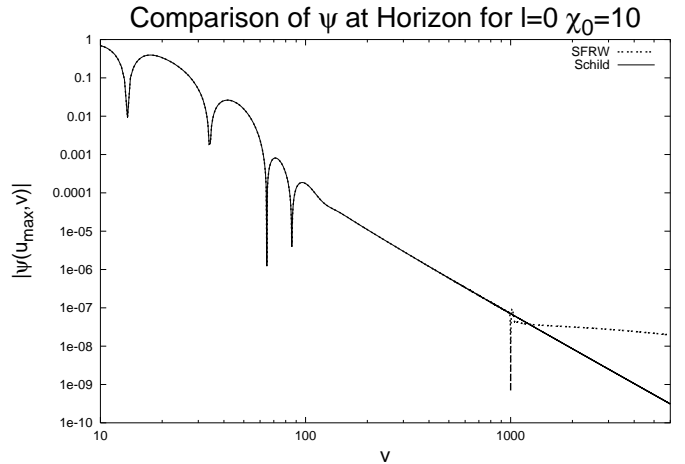}
\vspace*{2.6in}
\caption{Absolute value of the field on the event horizon as a
function of $v$, for $\ell=0$.  The solid line is the evolution of the
the scalar wave in pure Schwarzschild spacetime.  The dashed line is
the evolution in the Schwarzschild-Einstein-de Sitter spacetime.  The
sharp dip near $v=1000$ is where the field learns about the boundary
$\Sigma$ and changes sign.  Before this point, the field decays as
$\psi \sim v^{-3}$. After this point, the field decays as $\psi \sim
v^{-1}$.}
\label{fig:hor0}
\end{figure}

\begin{figure}
\special{hscale=35 vscale=35 hoffset=-25.0 voffset=10.0
         angle=-90.0 psfile=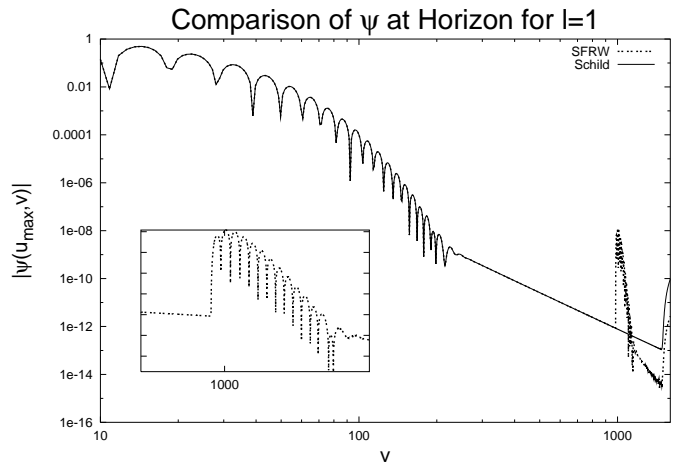}
\vspace*{2.6in}
\caption{Absolute value of the field on the event horizon, as a
function of $v$ for $\ell=1$.  The solid line is the evolution in pure
Schwarzschild spacetime.  The dashed line is the evolution in the
Schwarzschild-Einstein-de Sitter spacetime.  The inset is a close-up
of the region near $v=1000$ where the field learns about the boundary
$\Sigma$. Echoes of the quasi-normal oscillations in the field can be
seen.}
\label{fig:hor1}
\end{figure}

The numerical code evaluates the field on the event horizon, on the
boundary between the two regions of the spacetime, and at future null
infinity.  Our discussion here will be restricted to the value of the
field on the event horizon for the $\ell=0$ and $\ell=1$ modes. The
evolution was started at a ``late'' time, meaning that the boundary
has expanded far enough that it can be clearly seen that the field
initially behaves as it would in pure Schwarzschild spacetime.  For
both modes considered we see in figures \ref{fig:hor0} and
\ref{fig:hor1} that the field first exhibits quasi-normal ringing
followed by the well known power law decay (see, among others, 
\cite{2}). However, at a certain time in the evolution, the
field's behaviour deviates from the behaviour exhibited in the pure
Schwarzschild case.  The point at which the field changes behaviour
corresponds to the time at which information about the existence of
the cosmological region reaches the event horizon.

\begin{figure}
\special{hscale=50 vscale=50 hoffset=130.0 voffset=-200.0
         angle=45.0 psfile=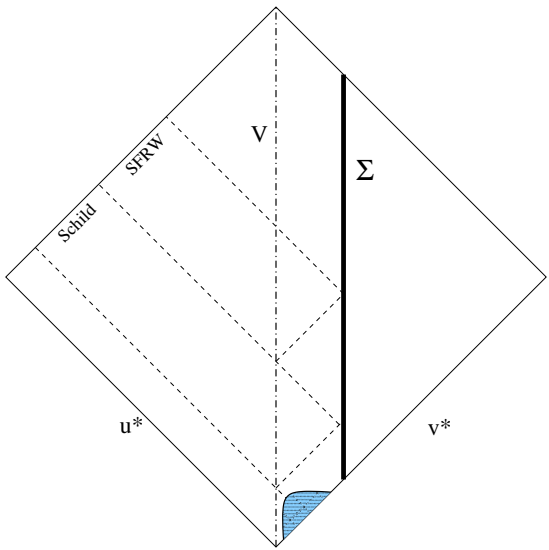}
\vspace*{2.6in}
\caption{Evolution of the scalar field.  Here the boundary between the
two spacetimes is the line $\Sigma$, with the black hole region to the
left and the cosmological region to the right. The line marked V
represents the maximum of the potential.  The dashed lines are the
reflection and transmission of the wave pulse from the potential and
the boundary. At the event horizon the field initially behaves as if
it were in pure Schwarzschild spacetime (Schild), then evolves
differently (SFRW).}
\label{fig:wave}
\end{figure}

As the wave packet falls towards the event horizon (approximated by
$u^*=u_{\rm max}$, where $u_{\rm max}$ is the largest value of $u^*$
in the numerical grid) it encounters the localized potential (dashed
line --- see Fig.~\ref{fig:wave}). Part of the wave is transmitted
through the barrier and reaches the event horizon, and part of the
wave is back-scattered by the potential.  The reflected wave heads out
towards the cosmological region (to the right), where it encounters
the boundary $\Sigma$.  For the $\ell=0$ mode the potential at the
boundary is discontinuous and negative, and the field now changes
sign.  This sign change is the large dip in Fig.~\ref{fig:hor0}
(note that this is a log scale, so that as the field passes through
$\psi=0$ the logarithm goes to negative infinity ).  It is when this
information reaches the event horizon that the evolution of the field
deviates from its evolution in pure Schwarzschild spacetime.  The
field continues to decay with a power-law falloff, but the falloff is
much slower than in the pure Schwarzschild case.

The discussion for the $\ell=1$ case is similar, until the reflected
wave reaches the boundary $\Sigma$.  This is because the potential at
the boundary is discontinuous and {\it positive} for $\ell>0$ (see
equation \ref{eq:VFRW}). Therefore the field will be partially
transmitted through the barrier at the boundary and partially
reflected off. The transmitted wave will make its way off to future
null infinity.  As for the part of the wave packet that has now been
reflected twice, it will fall back towards the black hole where it
once again encounters the localized potential.  The part of the wave
that manages to make it through the potential on its second encounter
heads back towards the event horizon, carrying information about the
existence of the boundary.  This second encounter with the localized
potential has the same effect on the packet as it did the first
time---namely, the field again exhibits quasi-normal ringing (see
inset of Fig.~\ref{fig:hor1}).  This ``echoing'' phenomenon occurs
only for the $\ell>0$ modes of the field.

\newpage

\section*{Acknowledgments}

The work presented in Part I was carried out in collaboration with
Patrick Brady and Chris Chambers; additional details can be found in
Ref.~\cite{15}. This work was supported by the Natural Sciences and
Engineering Research Council.

\end{document}